\documentstyle[preprint,aps,epsfig]{revtex}

\begin{document}

% Makes PACS numbers print 
\draft

\title{A renormalized expression for the turbulent energy dissipation rate} 
\author{C. Johnston and W. D. McComb\footnote{email:
W.D.McComb@ed.ac.uk}}
\address{Department of Physics and Astronomy, University of Edinburgh,
Edinburgh, EH9 3JZ}
\date{\today}
\maketitle

\begin{abstract}
Conditional elimination of degrees of freedom is shown to lead to an
exact expression for the rate of turbulent energy dissipation in terms
of a renormalized viscosity and a correction.  The correction is
neglected on the basis of a previous hypothesis \cite{McComb00} that
there is a range of parameters for which a quasi-stochastic estimate
is a good approximation to the exact conditional average.  This
hypothesis was tested by a perturbative calculation to second order in
the local Reynolds number, and the Kolmogorov prefactor (taken as a
measure of the renormalized dissipation rate) was
found to  reach a fixed point which was insensitive to initial values
of the kinematic viscosity and to values of the spatial rescaling factor
$h$ in the range $0.4\leq h\leq 0.8$. 
\end{abstract} 

% Insert PACS numbers
\pacs{47.27.Ak, 47.27.Eq, 47.27.Gs, 05.20.-y}

%body of the paper
\narrowtext
In the numerical simulation of fluid turbulence, as in other areas of
computational physics, there is a practical requirement to reduce the
number of degrees of freedom explicitly simulated.  However,
turbulence has the noteworthy requirement that any such reduction must
maintain the rate at which energy is dissipated.  In this letter we put
forward a method of renormalizing the energy dissipation rate.

It seems to be widely understood that any attempt to reduce the number
of degrees of freedom in the theoretical description of fluid turbulence
requires some form of conditional average, in which the retained
modes are kept unaveraged \cite{CArefs}. Yet this requirement has not
been recognised in most attempts to apply the dynamical
Renormalization Group (RG) algorithm to turbulence.  Normally such
methods rely instead on a band-filtered unconditional average.  A
critical appraisal of some of the leading approaches in this area will
be found in the paper by Eyink \cite{Eyink94}.

Originally, our own work, although introducing some features of the
conditional average, also relied on the use of the band-filtered
unconditional average \cite{McComb85}. Later it was recognized that a
conditional average in turbulence can only be carried out as an
approximation, and the two-field decomposition was introduced to
separate out random and deterministic effects
\cite{McComb92a,McComb92b}.  Recently, we have redefined the
conditional average in the form of a limit, eliminating the need to
separate into two fields \cite{McComb00}.  

The development reported here is that one of the corrections (to mode
elimination) in the momentum equation, vanishes identically in the
energy equation (and hence does not contribute to energy transfer),
while a second correction contributes to the energy spectrum but
vanishes identically in the equation for the dissipation rate.

We consider homogeneous, isotropic, incompressible, stationary
turbulence, with dissipation rate $\varepsilon$ given by
\begin{equation}
	\varepsilon = \int_0^\infty \text{d}k \, 2 \nu_0 k^2 E(k)
	\simeq \int_0^{k_0} \text{d}k \, 2 \nu_0 k^2 E(k),
\end{equation}
where the approximate equality defines the maximum wavenumber $k_0$. 
The value of $k_0$ is of the same order as the Kolmogorov dissipation
wavenumber $k_d^{(0)} = (\varepsilon / \nu_0^3 )^{1/4}$, where $\nu_0$
is the kinematic viscosity. 

In such turbulence the pair-correlation takes the
form $\left< u_\alpha (\bbox{k},t) u_\beta (\bbox{k'},s) \right> =
D_{\alpha\beta} (\bbox{k}) Q(\bbox{k};t,s) \delta(\bbox{k+k'})$,
where $D_{\alpha\beta}  (\bbox{k}) = \delta_{\alpha\beta} -
k_\alpha k_\beta / k^2$, and the energy spectrum is related to the
spectral density by $E(k,t) = 4 \pi k^2 Q(k;t,t)$.  To complete the
specification of our problem, we assume that energy is being injected
into some low range of wavenumbers by a source term $W(k)$, which
satisfies
\begin{equation}
	\int_0^{\kappa} \text{d}k \, W(k) = \varepsilon,
\end{equation}
for some $\kappa \ll k_d^{(0)}$.  This ensures stationarity.

Next we introduce a version of the RG which leads to a renormalized
dissipation rate equation.  The Navier-Stokes equation (NSE) may be
written in dimensionless form as:
\begin{eqnarray}
	\{ \partial_{\hat{t}} && + \hat{\nu}_0(\hat{k}) \hat{k}^2 \}
	\hat{u}_\alpha (\hat{\bbox{k}},\hat{t})= \nonumber \\
	&& = R_0 (k_0) M_{\alpha\beta\gamma}(\hat{\bbox{k}}) \int
	\text{d}^3\hat{j} \, \hat{u}_\beta(\hat{\bbox{j}},\hat{t})
	\hat{u}_\gamma (\hat{\bbox{k}}-\hat{\bbox{j}}, \hat{t}),
\end{eqnarray}
on $0 < \hat{k} < \hat{k}_0 = 1$, where
$\hat{k}=k/k_0$, $\hat{t}=t/\tau(k_0)$, $\hat{u}_\alpha
(\hat{\bbox{k}},t) = u_\alpha(\bbox{k},t)/V(k_0)$, $\tau(k_0)$ is
an, as yet, undefined timescale, $V(k_0)$ is the r.m.s.\ value of a
velocity mode with $|\bbox{k}|=k_0$, defined for any $k$ by
\begin{equation}
	V^2(k) = \frac{1}{k^3} \langle u_\alpha
	(\bbox{k},t) u_\alpha(-\bbox{k},t) \rangle,
\end{equation}
$R_0 (k_0) = \tau(k_0) V(k_0) k_0^4$ is the local Reynolds number (see
Batchelor \cite{Batchelor71}, p107) and
$	M_{\alpha\beta\gamma}(\bbox{k}) = (2i)^{-1} [ k_\beta
	D_{\alpha\gamma}(\bbox{k}) + k_\gamma
	D_{\alpha\beta}(\bbox{k}) ]$.
It should also be noted that the local Reynolds number is indeed
non-dimensional, since $u_\alpha (\mathbf{k},t)$ has dimensions
$L^4T^{-1}$.  The dynamical RG algorithm can now be stated as follows:
\begin{enumerate}
\item Rescale all wavevectors on $\hat{k}_1$ (=$h\hat{k}_0$), where $0 <
h < 1$, for example
$k'=\hat{k}/\hat{k}_1$, such that $ 0 < k' < h^{-1}$, and then average
out the effects of the high wavenumber modes to obtain a dynamical
equation for the modes on the interval $0 < k' < k'_1 \,(\equiv 1)$.
\item Use this low-wavenumber NSE to obtain the low-wavenumber energy
balance equation.
\item Integrate the energy balance equation with respect to $k'$ up to
the value $k'=k'_1\,(\equiv 1)$ to derive an equation for the
dissipation rate. 
\item Repeat these steps until the dissipation rate reaches a fixed 
point, at a new maximum wavenumber $k'_N=h^N k'_0$.
\end{enumerate}
The first step is to write the velocity fields as
\begin{equation}
	\hat{u}_\alpha^{\pm}(\hat{\bbox{k}},\hat{t}) =
	V(k_1)\psi_\alpha^{\pm} (\bbox{k'},t'), 
\end{equation}
where $\psi_\alpha^- (\bbox{k'},t')$ is defined on $0 < k' < 1$ and
$\psi_\alpha^+ (\bbox{k'},t')$ is defined on $ 1 < k' < h^{-1}$. The
combined (low-$k$ and high-$k$) equation of motion then takes the
form:
\begin{eqnarray}
	\{ && \partial_{t'} + \nu_0'(k') k'^2 \} \psi_\alpha^{\pm}
	(\bbox{k'},t') = \nonumber \\
	&& = R_1 (k_1) M_{\alpha\beta\gamma}^{\pm} (\mathbf{k'}) \int
	\text{d}^3 j' \, \psi_\beta (\bbox{j'},t') \psi_\gamma
	(\bbox{k'-j'},t'),
\end{eqnarray}
where $R_1 (k_1) = \tau(k_1) V(k_1) k_1^4$.
Next we average out the effect of the high-$k$ modes, while leaving
the low-$k$ modes unaffected.  In general this will require the
conditional projection of some functional $\mathcal F[\psi_\alpha]$ on
the  $\psi^-_\alpha$, which we denote by a subscript `$c$',
viz. $\langle \cdot \rangle_c$.  This should not be confused with the
usual ensemble average, as denoted by $\langle \cdot \rangle$.  An
important property of the conditional average is that the constraint
is lifted by a further \emph{unconditional} average \cite{Papoulis}.

Taking the low-$k$ equation, as given by (6), we conditionally average
both sides, and decompose the right-hand side according to (5), to
obtain:
\begin{eqnarray}
	&& \{ \partial_{t'}  + \nu_0'(k') k'^2 \} \psi_\alpha^-
	(\bbox{k'}) = \nonumber \\
	&& = R_1 (k_1) M_{\alpha\beta\gamma}^- (\bbox{k'}) \int
	\text{d}^3 j' \, \{ \langle \psi_\beta^- (\bbox{j'})
	\psi_\gamma^- (\bbox{k'-j'})\rangle_c +  \nonumber \\
	&& \: + \, 2 \langle \psi_\beta^- (\bbox{j'})
	\psi_\gamma^+ (\bbox{k'-j'})\rangle_c + \langle \psi_\beta^+
	(\bbox{j'}) \psi_\gamma^+ (\bbox{k'-j'}) \rangle_c \}.
\end{eqnarray}
We shall see presently that only the last term on the RHS of (7)
contributes to the energy dissipation rate.  In \cite{McComb00} we
gave a method for the approximate calculation of this conditional
average.  We shall refer to this approximation as the
\emph{quasi-stochastic estimate} (QSE) of the conditional average and
denote it by $\langle \cdot \rangle_{QSE}$.  Then, rearranging (7) and
adding and subtracting quantities to leave it unaffected, we may write
the low-wavenumber equation as:
\begin{eqnarray}
	( && \partial_{t'} + \nu_0'(k') k'^2 ) \psi_\alpha^-
	(\bbox{k'}) - \nonumber \\
	&& - \, R_1 (k_1) M_{\alpha\beta\gamma}^- (\bbox{k'}) \int
	\text{d}^3 j' \, \langle \psi_\beta^+ (\bbox{j'})
	\psi_\gamma^+ (\bbox{k'-j'}) \rangle_{QSE} \nonumber 
	\\
	&& = R_1 (k_1) M_{\alpha\beta\gamma}^- (\bbox{k'}) \int
	\text{d}^3 j' \, \psi_\beta^- (\bbox{j'}) \psi_\gamma^-
	(\bbox{k'-j'}) + \nonumber \\
	&& \qquad + \, S_\alpha^- (\bbox{k'}|k_1'),
\end{eqnarray}
where
\begin{eqnarray}
	&& S_\alpha^-  (\bbox{k'}|k_1')  =  R_1 (k_1)
	M^-_{\alpha\beta\gamma} (\bbox{k'}) \times \nonumber \\
	&& \times \int \! \text{d}^3j'
	\{ \underbrace{ \langle \psi_\beta^- (\bbox{j'})
	\psi_\gamma^- (\bbox{k'-j'})\rangle_c - \psi_\beta^-
	(\bbox{j'}) \psi_\gamma^- (\bbox{k'-j'})}_{S_1} +
	\nonumber \\
	&& + \, 2 \, \underbrace{\langle \psi_\beta^- (\bbox{j'})
	\psi_\gamma^+ (\bbox{k'-j'}) \rangle_c }_{S_2} + \nonumber
	\\
	&& + \, \underbrace{ \langle \psi_\beta^+ (\bbox{j'})
	\psi_\gamma^+ (\bbox{k'-j'}) \rangle_c - \langle \psi_\beta^+
	(\bbox{j'}) \psi_\gamma^+ (\bbox{k'-j'})\rangle_{QSE} }_{S_3}
	\}. \nonumber \\
\end{eqnarray}
We now wish to obtain the QSE, $\langle \psi^+_\beta (\bbox{j'},t')
\psi^+_\gamma (\bbox{k'}-\bbox{j'},t') \rangle_{QSE}$, which appears
on the LHS of (8).  Following the procedure in \cite{McComb00}, we
form an equation of motion for this quantity from (6).  This is solved
perturbatively as a power series in $R(k_1)$ and the band-filtered
moments of the $\psi^+$.  As a result, the low-$k$ equation, after
eliminating the first band of modes, takes the form
\begin{eqnarray}
	\{ && \partial_{t'}  + \nu_0'(k') k'^2 \}
	\psi_\alpha^- (\bbox{k'}) - \! \int \! \text{d}s' \,
	A_{\alpha\beta}(\bbox{k'},t'-s') \psi_\beta^-
	(\bbox{k'},s')\nonumber \\
	&& = R_1 (k_1) M_{\alpha\beta\gamma}^- (\bbox{k'}) \int
	\text{d}^3 j' \, \langle \psi_\beta^- (\bbox{j'})
	\psi_\gamma^- (\bbox{k'-j'})\rangle_c \nonumber \\
	&& \qquad + \, S_\alpha^- (\bbox{k'}|k_1'),
\end{eqnarray}
where 
\begin{eqnarray}
	A_{\alpha\beta} && (\bbox{k'},t'-s') =
	D_{\alpha\beta} (\bbox{k'}) R_1^2  (k_1) \times \nonumber \\
	&& \times [A^{(0)}(k',t'-s')+A^{(1)}(k',t'-s')R_1(k_1) +
	\nonumber \\
	&& \qquad + \, A^{(2)}(k',t'-s')R_1^2(k_1) + \ldots \;].
\end{eqnarray}
The coefficients $A^{(0)},\, A^{(1)},\, A^{(2)}, \ldots$ depend on the
moments of $\psi^+$ of order $2,\,3,\,4, \ldots$ respectively. It
should also be noted that the even-order coefficients are real and the
odd-order are imaginary, since the expansion is effectively in
powers of $M_{\alpha\beta\gamma}(\bbox{k})$, which is imaginary.

We now form the energy balance equation for the explicit scales $k\leq
k_1$, by multiplying each side of (10) through by  $\psi_\alpha^-
(\bbox{-k'},t')$ and averaging unconditionally. We then multiply
through by appropriate factors to restore the original unscaled
variables, in the process introducing the energy spectrum $E(k)$. We
also add $W(k)$, as specified in (2), with the result:
\begin{eqnarray}
	( && \partial_{t} + 2\nu_0(k) k^2 ) E(k) +
	2 \int \text{d}s\,A(k,t-s)E(k,s) = \nonumber \\
	&& = W(k) + T(k) + 8\pi k^2 V^2(k_1) \langle S_\alpha^-
	(\bbox{k}|k_1)\psi_\alpha^- (\bbox{-k}) \rangle,
\end{eqnarray}
where $A(k) = tr \, A_{\alpha\beta}(\bbox{k})$ and $T(k,t) = \int
\text{d}j \,\tilde{T} (k,j, |\bbox{k}-\bbox{j}|;t)$  is the
usual transfer spectrum, with wavenumbers in the interval $0\leq
k,\,j,\,|\bbox{k} -\bbox{j}| \leq k_1$.

Lastly, we may form an equation for the rate at which energy is
transferred through the modes of the system.  Integrating (12) with
respect to $k$, we obtain
\begin{eqnarray}
	2 && \int_0^{k_1} \text{d}k \left[\nu_0k^2+A(k)\right]E(k) =
	\varepsilon + \nonumber \\
	&& + \, 8\pi \int_0^{k_1} \text{d}k \, k^2 V^2(k_1) \langle
	S_\alpha^- (\bbox{k}|k_1) \psi_\alpha^-(\bbox{-k})
	\rangle.
\end{eqnarray}
Note that the integral over the transfer term vanishes identically due
to the antisymmetry of  $\tilde{T}(k,j,|\bbox{k}-\bbox{j}|;t)$ under
interchange of $\bbox{k}$ and $\bbox{j}$ (see \cite{Batchelor71}, p85).

At this stage, all three renormalized conservation equations (for
momentum, energy and dissipation rate) are exact.  Now consider the
effect of the term $S_{\alpha}^-$, divided into three parts, as shown in
(9), and begin with the energy equation.  The conditional average
behaves as a stochastic variable under a further \emph{unconditional}
average \cite{Papoulis}. Thus for the first term we have (schematically)
\begin{eqnarray}
	\langle S_1 \psi_\alpha^-  \rangle && \sim 
	\int \text{d}^3 j' \, \{ \langle \langle \psi_\beta^-
	(\bbox{j'}) \psi_\gamma^- (\bbox{k'-j'})\rangle_c
	\psi^-_\alpha (\bbox{-k'}) \rangle - \nonumber \\
	&& \qquad - \, \langle \psi_\beta^- (\bbox{j'})
	\psi_\gamma^- (\bbox{k'-j'}) \psi_\alpha^-
	(\bbox{-k'})\rangle \} = 0,
\end{eqnarray}
and so the contribution from $S_1$ vanishes identically in the energy
equation.

Now we turn to the dissipation equation: evidently we need only consider
$S_2$ and $S_3$. The first of these gives
\begin{eqnarray}
	&&\int \text{d}^3 k' \, \langle S_2 \psi_\alpha^-  \rangle
	\sim \nonumber \\
	&& \sim \int \text{d}^3 k' \!\! \int
	\text{d}^3 j' \, \langle \psi_\beta^- (\bbox{j'})
	\psi_\gamma^+ (\bbox{k'-j'}) \psi^-_\alpha
	(\bbox{-k'}) \rangle = 0, 
\end{eqnarray}
by antisymmetry under interchange of $\bbox{k'}$ and $\bbox{j'}$.
It should be noted that this property holds only because both
wavenumbers are on the same interval. This is not the case regarding
the contribution from $S_3$, which is
\begin{eqnarray}
	&& \int \text{d}^3 k' \, \langle S_3 \psi_\alpha^-
	\rangle \sim \nonumber \\
	&& \sim \int \text{d}^3 k' \!\! \int
	\text{d}^3 j' \, \{ \langle \psi_\beta^+ (\bbox{j'})
	\psi_\gamma^+ (\bbox{k'-j'}) \psi^-_\alpha
	(\bbox{-k'}) \rangle - \nonumber \\
	&& \qquad - \, \langle \psi_\beta^+
	(\bbox{j'}) \psi_\gamma^+ (\bbox{k'-j'})
	\psi^-_\alpha (\bbox{-k'})\rangle_{QSE} \}.
\end{eqnarray}
However, note that the two terms will cancel under any circumstances in
which the QSE is a good model for the exact conditional average.

Now, in order to perform an RG-style iteration, we truncate the
expansion for $A_{\alpha\beta}$, as given by (11), at lowest
order and rename
\begin{equation}
R_1^2(k_1)A^{(0)}(k',0) = \delta\nu_0(k')k'^2.
\end{equation}
Note that $A^{(1)}$ is imaginary and therefore cannot contribute to
the dissipation rate. This means that we effectively neglect terms of
order $R_1^4(k_1)$ and higher. To this level of approximation,  (10)
can be written as
\begin{eqnarray}
	( \partial_{t'} && + \nu'_1(k') k'^2 ) \psi_\alpha^-
	(\bbox{k'})  =  S^-_\alpha(\bbox{k'}|k'_1) + \nonumber \\
	&& + \, R_1 (k_1) M_{\alpha\beta\gamma}^-
	(\bbox{k'}) \int \text{d}^3 j' \psi_\beta^- (\bbox{j'})
	\psi_\gamma^- (\bbox{k'-j'}),
\end{eqnarray}
for $0 < k' < 1$. The renormalized viscosity is given by
\begin{equation}
	\nu'_1 (k') = \nu'_0 (k') + \delta \nu'_0(k'),
\end{equation}
and the equation for the increment is
\begin{eqnarray}
	\delta \nu'_0(k') = && R_1^2(k_1) \lim_{\ell' \to h^{-1}}
	\Delta_0(k') \left [ \frac{(k'^2/2) + j'^2 - k'j'\mu}{j'^2 -
	k'j'\mu} \right] \nonumber \\
	&& + \,  {\mathcal O}  \left( R_1^4(k_1) \right),
\end{eqnarray}
where $\mu$ is the cosine of the angle between $\bbox{k'}$ and
$\bbox{j'}$ and $\Delta_0(k')$ is the two-field form of the
increment\cite{McComb92b}, given by
\begin{eqnarray}
	\Delta_0(k')=\frac{1}{k'^2} \int \text{d}^3j'
	\frac{L(\bbox{k'},\bbox{j'})
	\hat{Q}^+(\ell')}{\nu'_0  (j') j'^2 + \nu'_0
	(\ell') \ell'^2}.
\end{eqnarray}
Here $\ell' = |\bbox{k'-j'}|$, $1 < k', \ell' < h^{-1}$
and  $L(\bbox{k'},\bbox{j'}) = -2 M^-_{\delta\beta\gamma}(\bbox{k'})
M^+_{\beta\delta\epsilon}(\bbox{j'})
D_{\epsilon\gamma}^+(\bbox{k'}-\bbox{j'})$.  Equation (20) for the
increment to the viscosity involves $\lim |\bbox{k'-j'}| \to h^{-1}$
(see \cite{McComb00}), and if this limit is evaluated by taking the
$\hat{Q}^+(|\bbox{k'-j'}|)$ as an expansion in Taylor series about
$k'=1$, then we make contact with the two-field version of McComb
and Watt \cite{McComb92b}, with spectral density given by
\begin{equation}
	\hat{Q}(l') = \frac{1}{k_1 V^2(k_1)} \left\{ h^{11/3} -
	\frac{11}{3} h^{14/3} (l' - h^{-1}) \right\}.
\end{equation}

It should be noted however that the factor in square brackets in (20)
is new.  This arises because we were able to improve on the Markovian
approximation used in the earlier calculations of McComb \emph{et al.}
\cite{McComb85,McComb92b}.  Details of this analysis will be given in
a subsequent paper.  The equations for any iteration labelled $n$ can
be found by induction, and numerical calculation shows that the
renormalized viscosity approaches a fixed point for some $n=N$, where
in practice $N=5$ or $6$, for most values of the spatial rescaling
factor. At the fixed point, (13) may be rearranged to give
\begin{eqnarray}
	\varepsilon = && 2 \int_0^{k_N} \text{d}k \nu_N(k)k^2 \, E(k)
	- \nonumber \\
	&& - \, 8 \pi \int_{k_{N-1}}^{k_N} \text{d}k k^2 V^2(k_N)
	\langle S_3\psi_\alpha^-(\bbox{-k})\rangle ,
\end{eqnarray}
where $\langle S_3\psi_\alpha^-(\bbox{-k})\rangle$ is shown
schematically in equation (16).

Let us now consider how to assess this work. We begin by noting that
the renormalized `viscous term' in (18) is not an observable, even
though it may be calculated to any order using (11).  This is
because $S^-_{\alpha}$ varies from realization to realization of the
explicit scales. In contrast, the renormalized viscosity in 
(12) would be (if we took the step given in (17)) an observable, as
all terms in the equation have been averaged.  However, the
contribution from $S_2$ is likely to prove important for energy
transfer and, as this has been omitted from our calculation, we shall
leave discussion of this to a fuller account, and concentrate on (23)
for the dissipation rate, as we know that $S_2$ does not contribute to
this equation.

In order to calculate the renormalized dissipation rate, we assume a
power-law form for the spectrum $E(k)=\alpha \varepsilon^{r}k^s$. Then,
the requirement that the renormalized viscosity (19) and its increment
(20) scale in the same way, along with the conservation requirement of
(23), yield $r=2/3$ and $s=-5/3$, along with an expression for the
Kolmogorov prefactor  $\alpha$ (see equation (92) in reference \cite{McComb92b}). 

The theoretical prediction of $\alpha$ can be taken as a measure of
the predicted dissipation rate, and in Figure 1 we show the result
of such calculations, with $\alpha$ iterating to a fixed point for
several different starting conditions at one value of the spatial
rescaling factor.  The fixed point corresponds to the upper end of the
inertial range and the value of $\alpha$ at the fixed point agrees
well with the result obtained from numerical simulations.  Figure 2
shows the fixed-point value of $\alpha$ for a range of spatial
rescaling factors.  It is of interest to note that the chain-dotted
curve to the left depicts an earlier version of the theory, in which
the conditional average was approximated by a band-filtered
average\cite{McComb85}.  The dashed line shows the result of working
out the limiting form of the viscosity (the stochastic part) with
scale separation\cite{McComb92b}, while invoking a Markovian
approximation. The continuous line shows the effect of including the
square brackets in (20) and gives rise to a prediction of $\alpha =
1.62 \pm 0.02$ for $0.2 < \eta < 0.6$, where the bandwidth $\eta =
1-h$.  Incidentally, it is perhaps worth remarking that the limit
$\eta \to 0$ (which one would expect in the microscopic case) does not
exist for macroscopic turbulence.  This is a consequence of an exact
symmetry of the NSE: local energy transfer vanishes when the
wavevector triad takes the form of an equilateral triangle.

Lastly, there is the question: how good is our perturbation
calculation?  The expansion, which we truncate, is in powers of
$\lambda=R^2_n(k)$, with integrals over moments of the $\psi^+$ (where
$|\psi^+|_{\text{rms}} \le 1$).  With the maximum value
$\lambda=0.16$, this is a small parameter, but possibly not small
enough for the truncation to qualify as a rational approximation
\cite{VanDyke2}.  Accordingly, we may have to rely on the properties
of the moment expansion. Certainly, the next step is to work out the
magnitude of the fourth-order term.  This is the subject of current
work.

Both authors acknowledge the support and facilities provided by the
Isaac Newton Institute.  We also thank Prof.\ M.E. Cates, G. Fullerton,
A. Hunter and A. Quinn for reading the manuscript and making numerous
helpful comments.  CJ acknowledges the financial support of the
Engineering and Physical Sciences Research Council.

\begin{figure}
\begin{center}
	\epsfig{figure=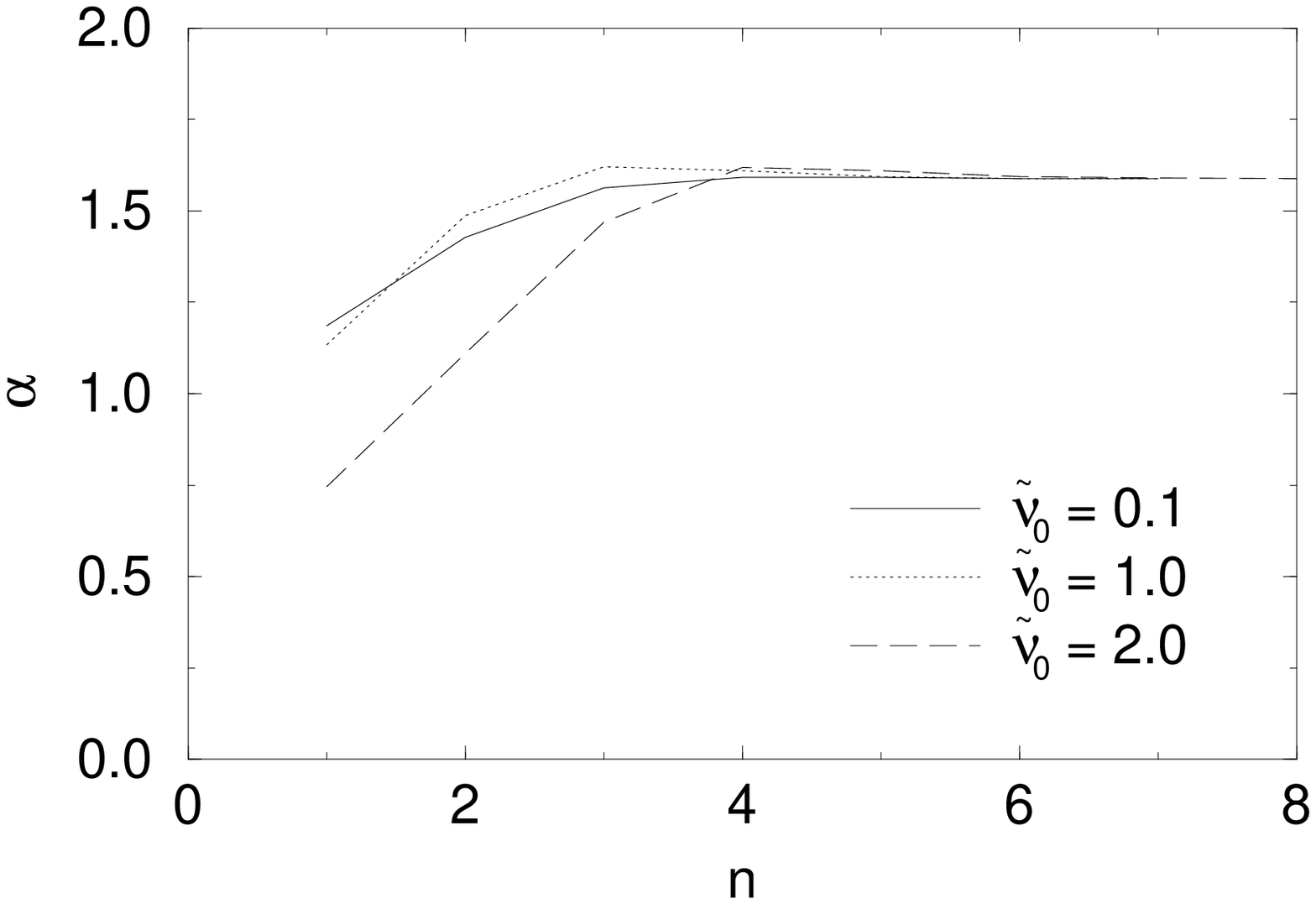, width = 80mm}
	\caption{\textsf{\small The Kolmogorov prefactor $\alpha$
	reaching a fixed point for a variety of starting
	viscosities $\tilde{\nu}_0$. (For the case where the spatial
	rescaling factor $h=0.60$ or the bandwidth $\eta=0.40$.)}} 
\end{center}
\end{figure}

\begin{figure}
\begin{center}
	\epsfig{figure=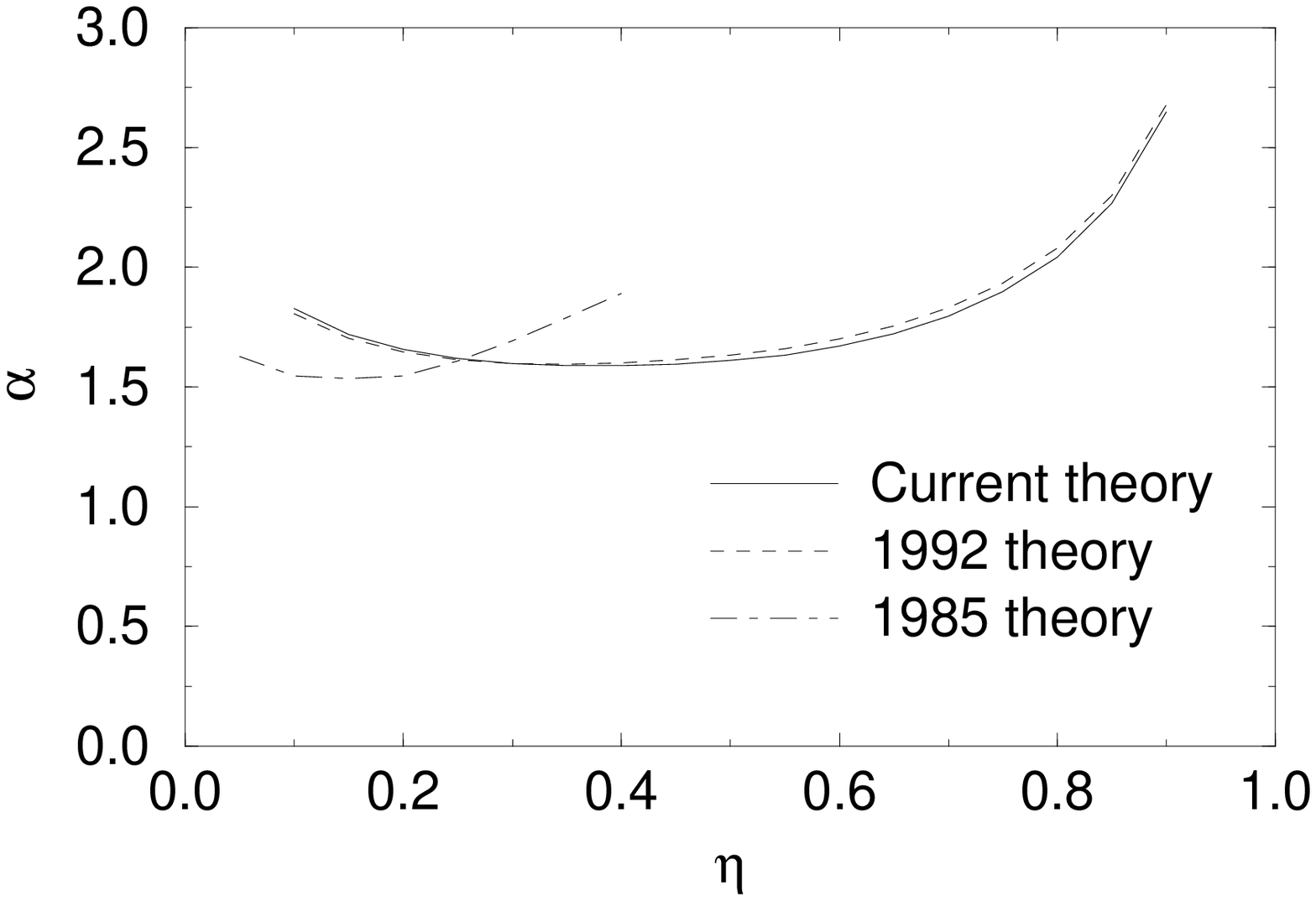, width = 80mm}
	\caption{\textsf{\small Variation of the Kolmogorov prefactor
	$\alpha$ with bandwidth $\eta$ or spatial rescaling factor
	$h$.}}
\end{center}
\end{figure}

\bibliographystyle{aip}

\end{document}